\documentclass[a4paper,fleqn,usenatbib]{mnras}

\usepackage[T1]{fontenc}
\usepackage{ae,aecompl}

\usepackage{graphicx}	
\usepackage{amsmath}	
\usepackage{amssymb}	
\usepackage{times}
\usepackage{graphics, subfigure}
\usepackage{epsfig}
\usepackage{multirow}
\usepackage{ulem}
\usepackage{pdfpages}
\usepackage{siunitx}

\def\simlt{\mathrel{\hbox{\rlap{\hbox{\lower4pt\hbox{$\sim$}}}\hbox{$<$}}}}
\def\simgt{\mathrel{\hbox{\rlap{\hbox{\lower4pt\hbox{$\sim$}}}\hbox{$>$}}}}

\newcommand{\mysim}{\mathord{\sim}}
\newcommand{\mylesssim}{\mathord{\lesssim}}

\newcommand{\myapprox}{\mathord{\approx}}

\newcommand{\logp}{\log P}

\title[Robust $H_0$]{A Cepheid systematics-free test of $H_0$ to $\lesssim2.5\%$ accuracy using SH0ES photometry}
\author[D. Kushnir and A. Sharon]{
Doron Kushnir$^{1}$\thanks{E-mail: doron.kushnir@weizmann.ac.il} and Amir Sharon$^{1}$
\\
$^{1}$Dept. of Particle Phys. \& Astrophys., Weizmann Institute of
Science, Rehovot 76100, Israel\\
}

\date{Accepted XXX. Received YYY; in original form ZZZ}

\pubyear{2020}

\begin{document}
\label{firstpage}
\pagerange{\pageref{firstpage}--\pageref{lastpage}}
\maketitle

\begin{abstract}
The recent SH0ES determination of the Hubble constant, $H_0=73.04\pm1.04$ km/s/Mpc, deviates significantly by $\myapprox5\sigma$ from the \textit{Planck} value, stimulating discussions on cosmological model extensions. To minimize statistical uncertainty and mitigate sensitivity to systematic errors in any single anchor distance determination, SH0ES combines Cepheids from various observations, including those from Type Ia supernova (SNe Ia) host galaxies, NGC 4258, and closer galaxies (MW, LMC, SMC, and M31), although this mixed sample may introduce unknown or subtle systematic errors due to comparing distant and closer Cepheids. To address this, we propose a subset excluding Cepheids from the closer galaxies, retaining only the NGC 4258 water megamasers as a single anchor, circumventing potential systematic errors associated with observational methods and reduction techniques. Focusing solely on these Cepheids yields competitive statistical errors, approximately $2.5\%$, sufficient to identify a $\myapprox3\sigma$ tension with the \textit{Planck} $H_0$ value. Our approach offers an opportunity to utilize optical photometry with systematic uncertainty smaller than the statistical uncertainty, potentially achieving higher precision than NIR photometry, given the lower optical background. However, currently the optical photometry sample's fidelity does not match that of NIR photometry. The significant Hubble tension obtained is unrelated to Cepheids and we discuss other options.
\end{abstract}

\begin{keywords}
cosmological parameters -- distance scale -- stars: variables: Cepheids-- supernovae: general
\end{keywords}



\section{Introduction}
\label{sec:Introduction}

The most recent determination of the Hubble constant by the SH0ES collaboration \citep[][hereafter R22]{Riess2022}, $H_0=73.04\pm1.04$ (in units of $\rm{km}\,\rm{s}^{-1}\,\rm{Mpc}^{-1}$ hereafter), exhibits a significant deviation of $\mysim5\sigma$ from the \textit{Planck} value \citep{Planck2020}, $H_0=67.4\pm0.5$, commonly referred to as the Hubble tension. This discrepancy between the Cepheid- and Type Ia supernovae (SNe Ia)-based SH0ES measurement and the cosmic microwave background temperature and polarization anisotropies \textit{Planck} measurement has spurred numerous proposals for extensions of the standard $\Lambda$CDM cosmology model \citep[see][for a review]{DiValentino2021}.

The absolute distance scale utilized by SH0ES is grounded on the period-luminosity relation of Cepheids \citep[$P-L$ relation;][]{Leavitt1912}, measured in the Hubble Space Telescope (HST) F160W filter (similar to the NIR $H$ band). These Cepheids are situated in 37 SNe Ia host galaxies (hereafter referred to as \textit{host Cepheids}) along with other anchor galaxies with absolute distance measurements. The Hubble tension manifests as a discrepancy of $\mysim0.1-0.2\,\textrm{mag}$ in the magnitudes of SH0ES Cepheids \citep{Riess2019b,Efstathiou2020}, wherein the host Cepheids appear brighter than the $\Lambda$CDM prediction.

The approach adopted by the SH0ES collaboration to determine $H_0$ involves combining the host Cepheids with a diverse large sample of Cepheids obtained from various observations. This strategy offers several advantages, including minimizing the statistical uncertainty in $H_0$ and utilizing multiple independent geometrical distance anchors, which reduces sensitivity to potential systematic errors in any single anchor distance determination. The additional Cepheids reside at the anchor galaxy NGC 4258 \citep[at a distance of $\myapprox7.5\,\textrm{Mpc}$, hereafter referred to as \textit{N4258 Cepheids}; absolute distance from VLBI observations of water megamasers orbiting its central supermassive black hole;][]{Humphreys2013,Reid2019}, as well at much closer galaxies such as the SMC and LMC  \citep[at $\myapprox50\,\textrm{kpc}$; absolute distance from double eclipsing binaries observed using long-baseline NIR interferometry;][]{Pietrzynski2019,Graczyk2020} and the Milky Way \citep[MW, at $\mysim1\,\textrm{kpc}$; absolute distance from Gaia EDR3 parallaxes;][]{Riess2021a}. Cepheids from M31 (at $\myapprox750\,\textrm{kpc}$) are also incorporated to better constrain the $P-L$ relation. However, the comparison of distant host Cepheids (typically at distances of $\myapprox15-50\,\textrm{Mpc}$ and M101 at a distance of $\myapprox6.5\,\textrm{Mpc}$) to Cepheids much closer to us involves a measurement range of $\geq$ 10 magnitudes which could in principle result in systematic errors between such measurements.

The measurement of host and N4258 Cepheids involves HST imaging of galaxies at multiple epochs using visual and NIR filters. This imaging data is utilized to identify and select Cepheids, determine their periods, extract their photometry at each epoch and filter, and amalgamate all epochs to derive a single mean magnitude value for each filter. These mean magnitudes are then employed to construct the Wesenheit index \citep{Madore1982}\footnote{We follow the convention that a single Cepheid magnitude $x$ is the magnitude of intensity mean, $\langle x \rangle$, and colors $(x-y)$ stand for $\langle x \rangle-\langle y \rangle$.}: 
\begin{eqnarray}\label{eq:Wesenheit H}
W_{H}=\textrm{F160W}-R_{H}^{VI}\left(\textrm{F555W}-\textrm{F814W}\right),
\end{eqnarray}
where F555W and F814W are the observed magnitudes in the corresponding HST filters (similar to the optical $V$ and $I$ band, respectively) and $R_{H}^{VI}\approx0.4$ is chosen to achieve extinction-independent measurements (for some specific extinction law, see a detailed discussion in Section~\ref{sec:R_Extinction}). On the other hand, the measurement of close Cepheids involves a distinct process \citep{Riess2019a,Riess2021a,Li2021}:
\begin{enumerate}
\item Close Cepheids are identified and selected based on archival observations.
\item These Cepheids, which are significantly brighter (by a factor of $10^{3}-10^{9}$) than host Cepheids, necessitate a different observing technique with HST, often involving a rapid spatial scan or DASH mode, along with a correction for count-rate nonlinearity.
\item Due to the narrow field of view of HST, a limited number of close Cepheids can be observed in one orbit, typically resulting in observations at a single epoch, which are then phase-corrected to mean light using archival (usually ground-based) observations.
\item The blending of host Cepheids with their environment (either chance superposition of Cepheids on crowded backgrounds or light from stars physically associated with Cepheids) due to the HST's angular resolution\footnote{ \ang[angle-symbol-over-decimal]{;;0.1} corresponds to $\myapprox7-25\,\textrm{pc}$ for the hosts (and $\myapprox3\,\textrm{pc}$ for M101).} requires a correction to photometry known as crowding correction, which is not applied to close Cepheids.
\end{enumerate}
The SH0ES collaboration extensively studied the impact of these differences on the comparison between close and host Cepheids and compensated for them through various corrections (R22).

\begin{figure}
\includegraphics[width=0.48\textwidth]{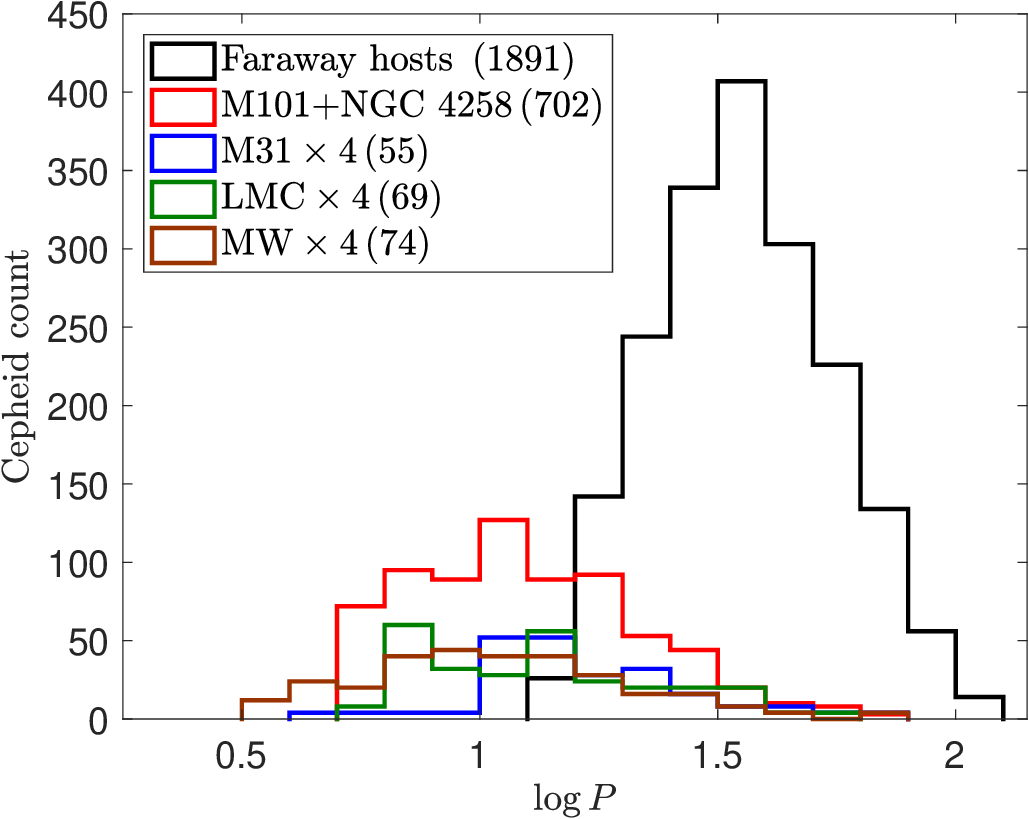}
\caption{The distribution of SH0ES Cepheid periods across different galaxies. Each $0.1$ $\log P$ bin showcases the count of Cepheids (multiplied by 4 in some cases for better visibility), with the total number displayed in the legend. Cepheids from various galaxy types are distinguished by color: Black represents faraway host galaxies, red indicates nearby galaxies (M101 and NGC 4258), blue denotes M31, green signifies the LMC, and brown represents MW Cepheids sourced from Table 1 in \citet{Riess2021a}. Notably, the majority of extragalactic (beyond M31) Cepheids exhibit $\log P>1$. Furthermore, Cepheid periods in nearby galaxies (M101 and NGC 4258) tend to be shorter on average compared to those in faraway galaxies, where the majority of Cepheids have $\log P>1.2$. 
\label{fig:Count}}
\end{figure}

Additionally, the large distance to host Cepheids biases their selection towards longer periods and lower extinction. Since long-period Cepheids are rarer, the close Cepheid sample is biased towards shorter period Cepheids, see Figure~\ref{fig:Count}. This discrepancy in period distributions is mitigated by assuming a specific shape of the $P-L$ relation, which may include breaks. Furthermore, the extinction of close Cepheids (in the MW and M31) tends to be higher than that of host Cepheids (see discussion in Section~\ref{sec:distribution}). To address this, observations are conducted in the NIR and the Wesenheit index is utilized. Finally, the metallicity of LMC and SMC Cepheids is lower than that of host Cepheids, which reside in galaxies specifically selected to have similar, roughly solar, metallicity. This difference is accounted for by assuming a metallicity-dependent $P-L$ relation.

The potential systematic errors arising from the aforementioned differences may introduce uncertainties as these effects lack a comprehensive theoretical understanding. Consequently, several studies have scrutinized the potential impact of such systematic errors on the determination of $H_0$ \citep{Follin2018,Anderson2018,Efstathiou2020,Riess2020,Javanmardi2021,Mortsell2021a,Mortsell2021b,Perivolaropoulos2021,Sharon2022}. However, none of these works have presented solid evidence of the existence of systematic biases. Nonetheless, it remains challenging to conclusively rule out a systematic bias (though see the comprehensive assessments in R22), thereby directing many investigations into potential systematic errors in the SH0ES measurement towards the Cepheids. In this paper, we propose a methodology to utilize the SH0ES Cepheids for $H_0$ measurement in a manner that is robust against Cepheid-related systematic errors. Our approach focuses solely on the hosts and N4258 Cepheids, capitalizing on the fact that the relative distances between the N4258 Cepheids and the hosts Cepheids are less susceptible to many systematic errors that may arise when comparing the hosts Cepheids to other anchor Cepheids. In essence, the NGC 4258 anchor stands out because it is the sole anchor galaxy for which the Cepheids are measured (both in terms of observation technique and data reduction) precisely as the hosts Cepheids.

For instance, consider the correction for blending photometry applied to the hosts Cepheids. Previous works found no evidence that this correction introduces systematic biases, both employing HST \citep[see R22 and the amplitude tests in][which are relevant for $\logp<1.7$\footnote{$\logp\equiv \log_{10}(P\,[\textrm{d}])$}]{Sharon2022} and JWST observations \citep[][who observed NGC 4258 and five hosts and found no significant difference in the mean distance measurements determined from HST and JWST]{Riess2024}. Nevertheless, it is challenging to rule out the possibility that a significant portion of long-period Cepheids (that are more massive and younger) are physically associated with stars. For example, \citet{Anderson2018} demonstrated by observing Cepheids in M31 that long period Cepheids have a higher chance of being in open clusters (see their Figure 13). The available data, however, are limited to Cepheid ages older than $\mysim50\,\rm{Myr}$ \citep[see also][with similar age limitations]{Breuval2023}. The ages of the long-period Cepheids are $\lesssim20\,\rm{Myr}$ \citep[see, e.g, Table A1 of][]{Anderson2016}, probably shorter than the dispersing time of open clusters. It is, therefore, reasonable to assume that a significant fraction of long-period Cepheids reside in open clusters. In such scenarios, the photometry of these long-period Cepheids would be biased, as the open clusters are not resolved for the hosts, and the crowding corrections applied by the SH0ES collaborations cannot rectify this effect. When comparing these Cepheids to the close Cepheids, where the cluster is resolved, their relative distances would be biased. While we do not claim the existence of such an effect \citep[and existing estimates suggest it is small;][]{Anderson2018}, it inherently cancels out when comparing the hosts Cepheids to the N4258 Cepheids, as the open clusters affect both in the same manner. Other potential systematic errors, such as those in identification, selection, period determination, photometry, and mean magnitude determination, are likewise eliminated or minimized. Therefore, we propose utilizing a subsample of SH0ES Cepheids that excludes the MW, LMC, SMC, and M31 Cepheids, effectively making the determination of $H_0$ impervious to Cepheid-related systematic errors.

By excluding the MW, LMC, SMC, and M31 Cepheids, the majority of Cepheids in the sample exhibit $\logp>1$ (see Figure~\ref{fig:Count}). Consequently, all Cepheids with $\logp<1$ can be disregarded, effectively mitigating the sensitivity to breaks in the $P-L$ relation. Moreover, Cepheids in the remaining sample exhibit similar metallicities, and the extinction towards them is minimal. Although systematic differences persist between M101 and NGC 4258 (hereafter \textit{nearby galaxies}) and other (hereafter \textit{faraway}) galaxies, such as the average shorter periods of Cepheids in nearby galaxies compared to those in faraway galaxies (see Figure~\ref{fig:Count}), we demonstrate in Section~\ref{sec:WH results} that these systematic differences have negligible effects on the determination of $H_0$. With this refined sample, the remaining statistical errors in determining the Cepheid relative distance scale, specifically those associated with the zero point of the Cepheid brightness (attributed to intrinsic scatter, measurement error, and finite number of Cepheids) and the SNe Ia brightness (due to intrinsic scatter and measurement error), are sufficiently small to be competitive (Section~\ref{sec:WH results}). The statistical error is approximately $\myapprox2.3\%$ (or about 1.65 km/s/Mpc), which is adequate to identify a $\myapprox3\sigma$ tension with the \textit{Planck} $H_0$ value.

The systematic errors associated with Cepheids in our approach are significantly smaller than the statistical uncertainty. Specifically, we illustrate that uncertainties related to the reddening law and its free parameters, metallicity sensitivity, and breaks in the $P-L$ relation (or even the existence of a global $P-L$ relation) are either eliminated or substantially reduced. Since these sensitivities are the rationale for opting for NIR over optical photometry, there exists an opportunity to utilize optical photometry with significantly reduced systematic errors stemming from these effects. In this case, the relevant Wesenheit index becomes:
\begin{eqnarray}\label{eq:Wesenheit I}
W_{I}=\textrm{F814W}-R_{I}^{VI}\left(\textrm{F555W}-\textrm{F814W}\right),
\end{eqnarray}
where $R_{I}^{VI}\approx1.3$ is chosen for $W_I$ to be independent of extinction \citep[e.g.,][]{Riess2019a}. In Section~\ref{sec:WI results}, we demonstrate that the systematic uncertainty with optical photometry is markedly smaller than the statistical uncertainty, potentially allowing for higher precision than that achieved with NIR photometry, as the optical background is nearly an order of magnitude lower than that in the NIR, owing to higher resolution, smaller pixels, and lower flux from red giants (R22). The reduced sensitivity to extinction correction is attributed to the minimal extinction of Cepheids in our sample (Section~\ref{sec:Low extinction}). Upon scrutinizing the optical optical datasets for the 19 hosts of \citet[][hereafter R16]{Riess2016}, we find that, in practice, the photometry of the optical bands is not much improved compared to the NIR photometry and the fidelity of the optical photometry sample is not at the level of the NIR photometry sample (see Section~\ref{sec:WI results}). 

The key advantage of our approach is that the significant Hubble tension obtained is unrelated to Cepheids. Given the absence of any indications of SNe Ia-related systematic errors following detailed scrutiny \citep[R22,][but see also \citealt{Steinhardt2020,Wojtak2022,Wojtak2024}]{Carr2022,Brout2022,Peterson2022}, our analysis suggests that either the standard cosmology is flawed or there exists a systematic error in the current distance determination to NGC 4258. We delve into this matter and summarize our findings in Section~\ref{sec:summary}.


\section{Fitting method and data}
\label{sec:fit method data}

In this section, we describe our fitting technique (subsection~\ref{sec:fit method}) and the dataset (subsection~\ref{sec:data}) used for our main fits. These fits specifically exclude Cepheids from the MW, LMC, SMC and M31, along with those with periods $\logp<1$.

\subsection{Fitting method}
\label{sec:fit method}

We adopt a fitting methodology similar to that of \citet{Mortsell2021a}, which bears resemblance to the approach outlined in R16. We express the Wesenheit magnitude, elaborated upon in Section~\ref{sec:R_Extinction}, in the $k$ band (either $I$ for F814W or $H$ for F160W) of the $j$th Cepheid within the $i$th galaxy as follows:
\begin{eqnarray}\label{eq:Cepheid fit}
m_{k,i,j}^{W}&=\mu_{i}+M_{k}^{W}+b^{W}_{k}\left[P\right]_{i,j}+Z^{W}_{k}\Delta\log_{10}\left(\rm{O}/\rm{H}\right)_{i,j}.
\end{eqnarray}
Here, $\left[P\right]_{i,j}=\log P_{i,j}-1$, $M_{k}^{W}$ denotes the absolute Cepheid magnitude normalized to a period of $\logp=1$ and a fiducial metallicity of $\log_{10}\left(\rm{O}/\rm{H}\right)=8.9$, $\Delta\log_{10}\left(\rm{O}/\rm{H}\right)_{i,j}=\log_{10}\left(\rm{O}/\rm{H}\right)_{i,j}-\log_{10}\left(\rm{O}/\rm{H}\right)$ represents the difference between the estimated Cepheid metallicity and the fiducial value, and $\mu_i=5\log_{10}D_{i}+25$ stands for the distance modulus to the $i$th galaxy, with $D_i$ being the luminosity distance in Mpc. The nuisance parameters $b^{W}_{k}$ and $Z^{W}_{k}$ define the relationship between Cepheid period, metallicity, and luminosity. An alternative analysis employing period bins is conducted, where for each Cepheid in the $l$ period bin, we express:
\begin{eqnarray}\label{eq:Cepheid fit bin}
m_{k,i,j}^{W}&=\mu_{i}+M_{k}^{W,l}+b^{W,l}_{k}\left[P\right]_{i,j}+Z^{W,l}_{k}\Delta\log_{10}\left(\rm{O}/\rm{H}\right)_{i,j}.
\end{eqnarray}
Here, $M_{k}^{W,l}$, $b^{W,l}_{k}$ and $Z^{W,l}_{k}$ define the relation in the $l$ period bin. An analysis incorporating a break in the $P-L$ relation entails two period bins with $M_{k}^{W,1}=M_{k}^{W,2}$ and $Z_{k}^{W,1}=Z_{k}^{W,2}$.

The magnitudes of SNe Ia in the calibration sample are expressed as:
\begin{eqnarray}\label{eq:SN fit}
m_{B,i,j}=\mu_{i}+M_{B},
\end{eqnarray}
Where $m_{B,i,j}$ represents the maximum-light apparent $B$-band brightness of the $j$th SNe Ia in the $i$th host\footnote{or, possibly, to the $j$th measurement of the same SN Ia.} at the time of $B$-band peak, corrected to the fiducial color and luminosity. The distance to NGC 4258, $\mu_{\rm{N}2458}=29.397\pm0.032\,\rm{mag}$ \citep{Reid2019}, is incorporated as an additional data point.

We simultaneously fit for $b^{W}_{k}$, $Z^{W}_{k}$, $M_{k}^{W}$ (or for $b^{W,l}_{k}$, $Z^{W,l}_{k}$, $M_{k}^{W,l}$ in the case of the binned analysis), $M_{B}$, and the galaxy distances $\mu_i$. As the system of equations is linear, the fit can be conducted analytically \citep[see Appendix A of][]{Mortsell2021a}. We employ a global outlier rejection threshold of $2.7\sigma$. Our findings in Sections~\ref{sec:WH results}-\ref{sec:WI results} demonstrate that our results are relatively insensitive to the method of outlier rejection. Given $M_B$, we determine $H_0$ through:
\begin{eqnarray}\label{eq:H0 fit}
H_0=10^{M_{B}/5+a_B+5},
\end{eqnarray}
where $a_B=0.71273\pm0.00176\,\rm{mag}$ represents the intercept of the SNe Ia magnitude-redshift relation (R16).

\subsection{Cepheid Data}
\label{sec:data}

Since the latest extragalactic Cepheid optical dataset from R22 has not yet been made publicly accessible, we perform our analysis using two separate datasets, as made available through the data release of R22 \footnote{\url{https://github.com/PantheonPlusSH0ES/DataRelease}. The tables exclusively include Cepheids that pass the $3.3\sigma$ outlier rejection criterion of R22.}: 

\begin{itemize}
\item R22-$W_H$ -- This dataset incorporates $m_{H,i,j}^{W}$, $\left[P\right]_{i,j}$, $\log_{10}\left(\rm{O}/\rm{H}\right)_{i,j}$, and $\sigma_{H,i,j}^{W}$ (representing the total statistical uncertainty in $m_{H,i,j}^{W}$; we use the provided covariance matrix as well) for the full ($37$ hosts; $42$ SNe Ia) sample.
\item R22-$W_I$ -- Here, we utilize $m_{I,i,j}^{W}$, $\left[P\right]_{i,j}$, $\log_{10}\left(\rm{O}/\rm{H}\right)_{i,j}$, and $\sigma_{I,i,j}^{W}$ (covariance matric is not provided in this case) for the $19$ R16 hosts (and $19$ SNe Ia; hosts distances $\mylesssim30\,\textrm{Mpc}$). We scale $\sigma_{I,i,j}^{W}$ by a factor of $0.85$, as recommended in the data release notes (see the discussion in Section~\ref{sec:WI results}).
\end{itemize}

The R22 datasets incorporate several improvements compared to earlier data releases, as outlined in Section 3.4 of R22. R22 data release includes $m_{B,i,j}$ along with its corresponding covariance matrix.


\section{$W_H$ fitting results}
\label{sec:WH results}

\begin{figure*}
\includegraphics[width=1\textwidth]{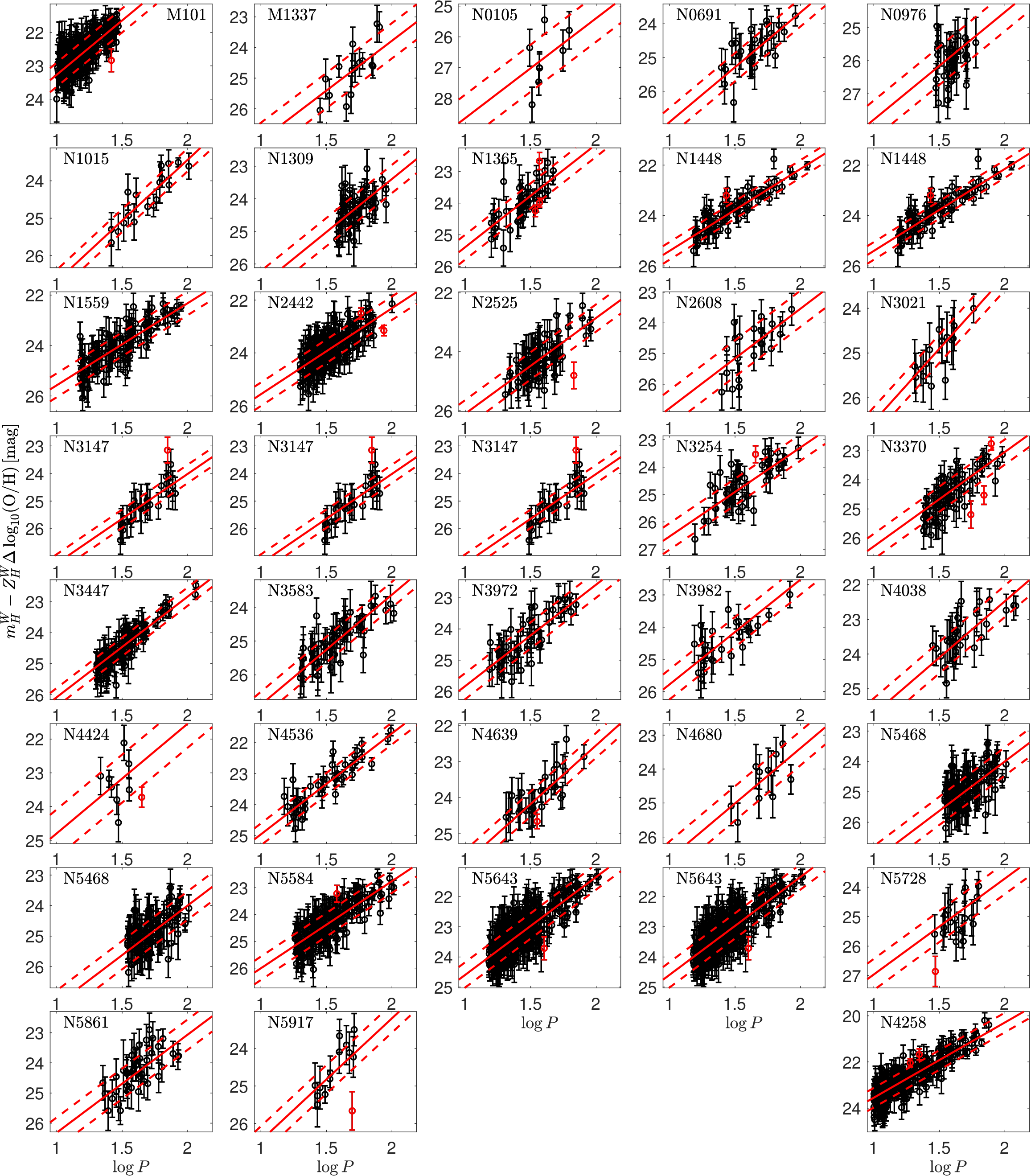}
\caption{The individual Cepheid $P-L$ relations derived from the primary $W_H$ fit. In each panel, the solid red line indicates the best fit, while the dashed red lines represent one standard deviation of the residuals within each galaxy around the optimal fit. Black symbols denote Cepheids that passed the outlier rejection, whereas red symbols represent those that did not.
\label{fig:LP}}
\end{figure*}

For our primary fit using the R22-$W_H$ dataset, we adopt $R_H^{VI}=0.386$ \citep[corresponding to $R_V=3.3$ of][for consistency with R22; see Section~\ref{sec:R_Extinction}]{Fitzpatrick1999}, resulting in $H_0=72.68\pm1.67$ (refer to case 1 in Table~\ref{tbl:H0} for detailed information). Figure~\ref{fig:LP} illustrates the individual Cepheid $P-L$ relations, where the solid red line indicates the best fit, and the dashed red lines represent one standard deviation of the residuals around the best fit within each galaxy. Cepheids failing the global $2.7\sigma$ outlier rejection criterion are highlighted in red. Our outcome closely aligns with the result of fit 10 from R22, which yielded $H_0=72.51\pm1.54$, utilizing NGC 4258 as the sole anchor while incorporating all ($\mysim3500$) Cepheids in the analysis.

The obtained error of approximately $\myapprox2.3\%$ in $H_0$ is linked to the error in $M_B$ through $\myapprox\delta M_{B}\ln(10)/5$, where 
\begin{eqnarray}\label{eq:dMB}
\delta M_B^2\approx\delta\mu_{\rm{N}4258,\rm{anc}}^2+\delta\Delta\mu_{\rm{N}4258}^2+\delta(\Delta\mu_{\rm{host}}+m_B)^2.
\end{eqnarray}
Here, $\delta\mu_{\rm{N}4258,\rm{anc}}\approx0.032\,\rm{mag}$ denotes the absolute distance error to NGC 4258 from megamaser observations \citep{Reid2019}. $\delta\Delta\mu_{\rm{N}4258}$ represents the NGC 4258 contribution to the error of the relative distance between NGC 4258 and the host galaxies based on the Cepheids $P-L$ relation. $\delta(\Delta\mu_{\rm{host}}+m_B)$ signifies the weighted error in the mean of each host's contribution to the relative distance and $m_{B}$. For each galaxy, we can estimate $\delta\Delta\mu_i\approx(\sum_{m,n} C^{-1}_{mn})^{-1/2}$, where $C$ is the covariance matrix for $\sigma_{H,i,j}^{W}$, and $\delta m_{B,i}\approx(\sum_{m,n} D^{-1}_{mn})^{-1/2}$, where $D$ is the covariance matrix for $\delta m_{B,i,j}$. This estimate leads to $\delta\Delta\mu_{\rm{N}4258}\approx0.028\,\rm{mag}$, which can be approximated by $\delta\Delta\mu_{\rm{N}4258}\approx0.42/\sqrt{223}\approx0.028\,\rm{mag}$, where $\myapprox0.42\,\rm{mag}$ is the mean statistical uncertainty of the 223 N4258 Cepgeids' Wesenheit magnitude. Similarly, we obtain $\delta(\Delta\mu_{\rm{host}}+m_B)=[\sum(\delta \Delta\mu_i^2+\delta m_{B,i}^2)^{-1}]^{-1/2}\approx0.021\,\rm{mag}$, where the SNe Ia contribution to the error budget, $\delta_{m_B}=[\sum\delta m_{B,i}^{-2}]^{-1/2}\approx0.016\,\rm{mag}$, can be approximated by the $\myapprox0.11\,\rm{mag}$ SNe Ia intrinsic scatter, $\delta_{m_B}\approx0.11/\sqrt{37}\approx0.018\,\rm{mag}$. Our estimate yields $\delta M_B\approx0.047\,\rm{mag}$, resulting in an $\myapprox2.2\%$ error in $H_0$, consistent with the fit results.

\begin{table*}
\begin{minipage}{160mm}
\caption{Summary of fitting results for different cases. The Wesenheit index ($W_I$ or $W_H$) is presented in the second column. For $W_H$ we use the R22-$W_H$ dataset ($37$ hosts; $42$ SNe Ia; full covariance matrix for Cepheid Wesenheit index) and for $W_I$ we use the R22-$W_I$ dataset ($19$ hosts; $19$ SNe Ia; covariance matrix for Cepheid Wesenheit index was not provided). Anchor galaxies are specified in the third column (N for NGC 4258 and M for MW). The number of Cepheids included in the fit (after outlier rejection) is listed in the fifth column. The reduced chi-squared value and the corresponding degrees of freedom are provided in the eighth and ninth columns, respectively.}
\begin{tabular}{cccccccccc}
\hline
Case & Wesenheit  & Anc &      $H_0$ & $N$ & $Z^W_k$  & $b^W_k$              & $\chi^{2}_{\nu}$ & dof & comments  \\ 
         &   index  &    & $[\rm{km}\,\rm{s}^{-1}\,\rm{Mpc}^{-1}]$  &    & $[\rm{mag}\,\rm{dex}^{-1}]$ &  $[\rm{mag}\,\rm{dex}^{-1}]$ &    &   \\ 
  \hline
   & &  &   \multicolumn{3}{c}{$W_H$ fits, Section~\ref{sec:WH results}} & & &   \\
1 & $W_H$ & N & $72.68\pm1.67$ & $2306$ & $-0.07\pm0.12$ & $-3.22\pm0.05$ & $0.96$ & 2342 & primary \\
2 & $W_H$ & N & $72.50\pm1.69$ & $2302$ &                           &                           & $0.97$ & 2311 & $\logp$ bins \\
\hline
   & &  &   \multicolumn{3}{c}{$W_I$ fits, Section~\ref{sec:WI results}} & & &   \\
3 & $W_I$ & N & $76.16\pm1.69$ & $1285$ & $-0.43\pm0.12$ & $-3.20\pm0.04$  & $0.90$ & 1299 & primary \\
4 & $W_I$ & N & $75.25\pm1.71$ & $1275$ &                           &                             & $0.88$ & 1262 & $\logp$ bins \\ \hline
& &  &   \multicolumn{3}{c}{Section~\ref{sec:distribution}} & & &  \\
5 & $W_H$ & N & $72.39\pm1.66$ & $2312$ & $0.05\pm0.12$ & $-3.15\pm0.05$ & $0.92$ & 2348 & $R_{H}^{VI}=0$ \\ 
6 & $W_I$ & N & $74.42\pm1.66$ & $1290$ & $-0.11\pm0.11$ & $-2.84\pm0.04$ & $0.91$ & 1304 & $R_{I}^{VI}=0$ \\
\hline
& &  &   \multicolumn{3}{c}{Section~\ref{sec:MW}} & & &  \\
7 & $W_H$ & M & $74.43\pm2.21$ & $2632$ & $-0.17\pm0.09$ & $-3.25\pm0.03$ & $0.96$ & 2666 & \\ 
8 & $W_I$ & M & $75.67\pm2.24$ &  $1561$ & $-0.28\pm0.09$ & $-3.16\pm0.03$ & $0.88$ &  1573 &\\
9 & $W_H$ & M & $74.28\pm2.21$ &  $2628$ & $-0.01\pm0.09$ & $-3.14\pm0.03$ & $0.92$ & 2662  & $R_{H}^{VI}=0$ \\ 
10 & $W_I$ & M & $83.10\pm2.60$ &  $1547$ & $-0.05\pm0.09$ & $-2.76\pm0.03$ & $0.89$ & 1559 & $R_{I}^{VI}=0$ \\
\hline
\end{tabular}
\centering
\label{tbl:H0}
\end{minipage}
\end{table*}

We proceed to demonstrate that the systematic error associated with determining $H_0$ using our primary R22-$W_H$ fit is considerably smaller than the statistical error. Following the methodology of R22, we evaluate the systematic uncertainty by investigating the impact of various fit variants on the derived value of $H_0$. Most of the examined variants exhibit a negligible effect on $H_0$ ($|\delta H_0|<0.6$). These include variants such as incorporating Cepheids with $\logp<1$ (with or without a different $P-L$ slope; the effect is minimal due to the limited number of Cepheids in this range, approximately 250), exclusion of the outlier rejection, adjustment of the outlier rejection to eliminate the single largest outlier at a time, raising the outlier rejection threshold to $3.5\sigma$ (with either global or single-outlier-at-a-time rejection), adjusting the period cutoff to $\logp>1.2$ (which aligns the period distributions of nearby and faraway galaxies more closely), disregarding the metallicity term (setting $Z_H^{W}=0$), adopting $R_H^{VI}=0.3$ \citep[smaller than $R_V=2.5$ of][]{Fitzpatrick1999}, employing $R_H^{VI}=0.5$ \citep[larger than $R_V=3.3$ of][]{Cardelli1989}, and including M31 Cepheids (the effect is minimal due to the limited number of Cepheids in M31 with $\logp>1$, approximately 50). The sole variant exhibiting a non-negligible effect on the value of $H_0$ is the adjustment of the period cutoff to $\logp<1.6$ ($\delta H_0\approx-0.88$). Given the minimal impact of all fit variants, we can safely disregard the systematic uncertainty in this case.

\begin{figure}
\includegraphics[width=0.48\textwidth]{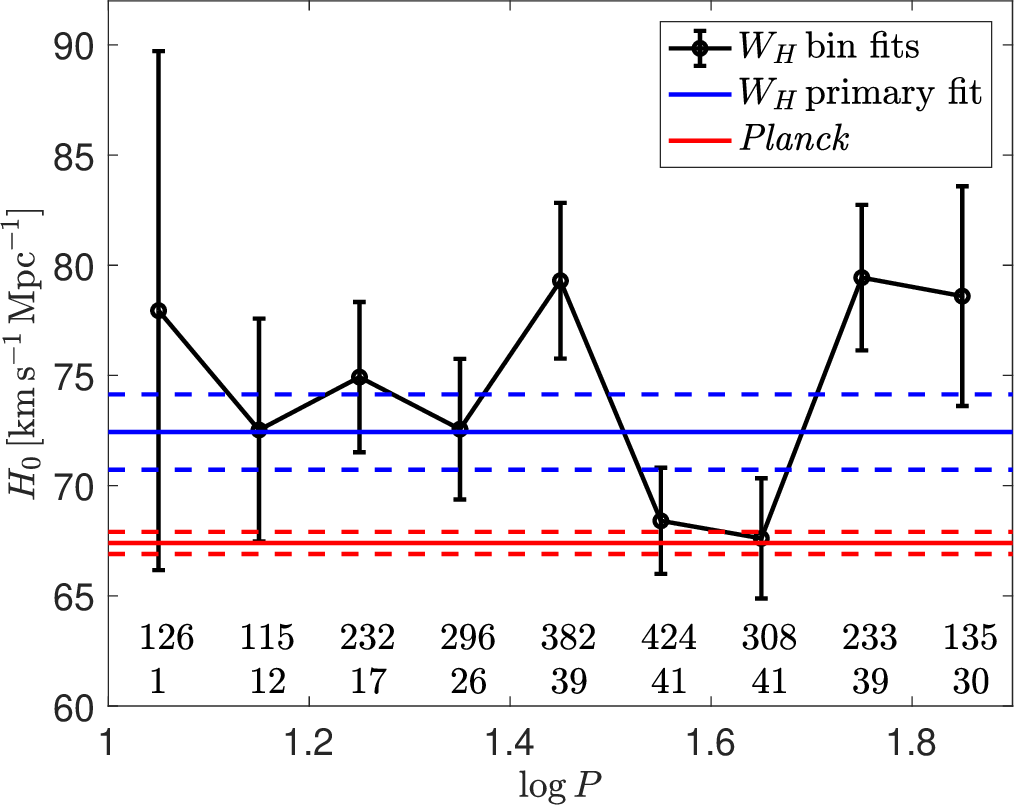}
\caption{Results of the fitting procedure using $W_H$ in $0.1-\logp$ bins. The fits' outcomes (depicted in black; errors denote one standard deviation) are dispersed around the primary fit result (shown in blue, where the solid line represents the best fit and dashed lines indicate one standard deviation), without exhibiting a clear dependence on period, and all align with the primary fit result (though the precision within each period bin is constrained by the quantity of Cepheids and SNe Ia, indicated for each bin). The red lines denote the \textit{Planck} results.
\label{fig:Bin_Plot}}
\end{figure}

\begin{figure*}
\includegraphics[width=1\textwidth]{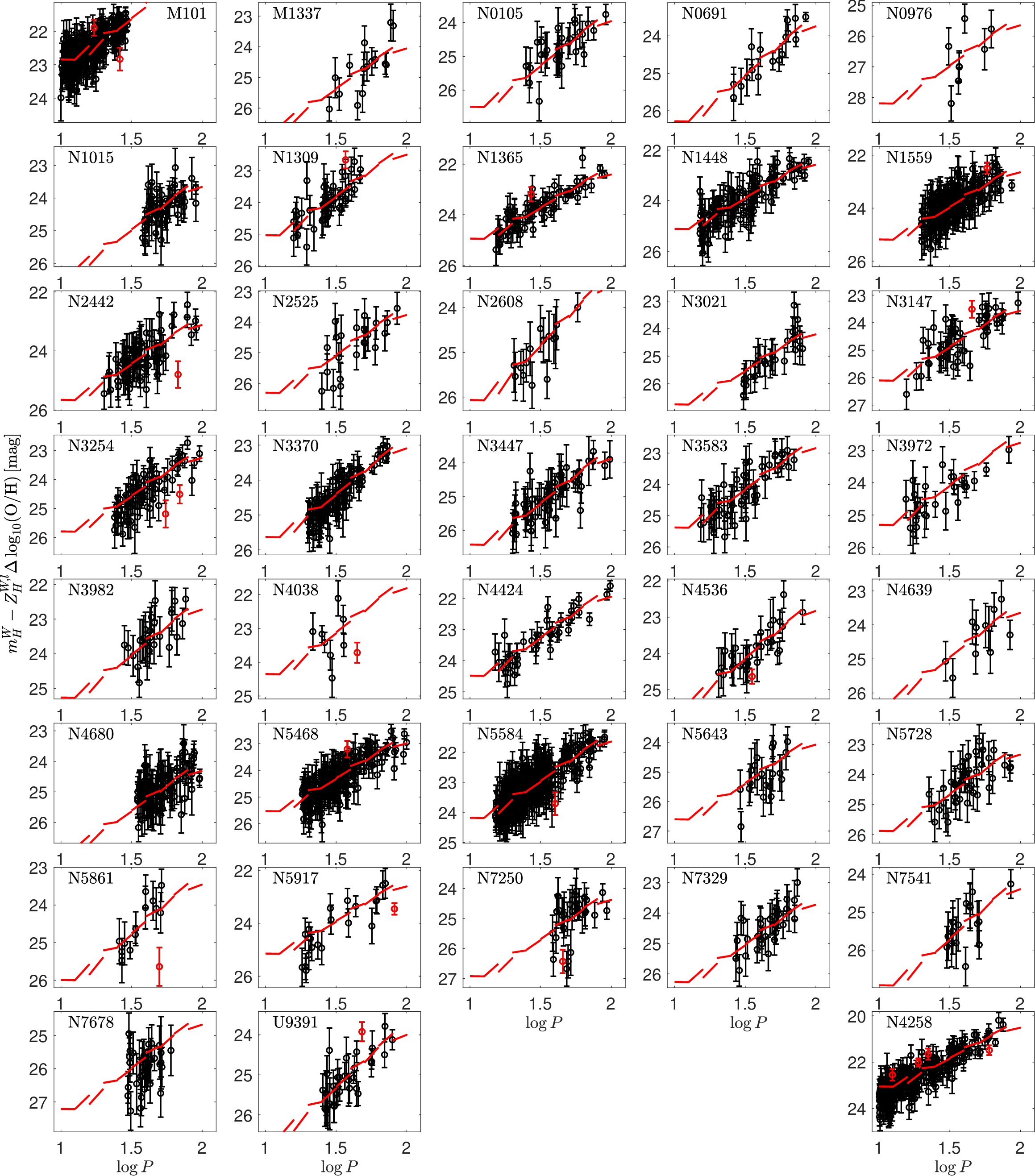}
\caption{The individual $W_H$ Cepheid $P-L$ relations for the simultaneous fit in $0.1-\logp$ bins. Each panel shows the best fit (in each period bin) represented by the broken solid red line. Black (red) symbols denote Cepheids that passed (did not pass) outlier rejection for the respective galaxy under examination. This fit remains unbiased by the varying distributions of periods in nearby and faraway galaxies. The derived value of $H_0$ is $72.09\pm1.75$, fully consistent with the primary fit result.
\label{fig:LP_bin}}
\end{figure*}

The robustness of our primary R22-$W_H$ fit stems from the uniformity of the Cepheid dataset utilized. The consistent observational methodology employed to determine photometry across this sample mitigates many potential systematic errors associated with photometry determination. Generally, any systematic biases that are power-laws of the period are absorbed within the $P-L$ relation, rendering them inconsequential. However, other functional forms or other systematic differences between nearby and faraway galaxies may still exert some influence. For instance, the average periods of Cepheids in nearby galaxies tend to be shorter compared to those in faraway galaxies (refer to Figure~\ref{fig:Count}). To scrutinize such effects rigorously, we propose a robust, overarching test. This test directly compares Cepheids with identical periods across nearby and faraway galaxies, without presupposing a universal $P-L$ relation. Before applying this test, it is insightful to revisit our primary fit within a narrow period bin. The outcomes of such fits within $0.1$-wide $\logp$ bins are depicted in Figure~\ref{fig:Bin_Plot}. Evidently, the fit results are distributed around the primary fit outcome, exhibiting no discernible period dependence, and all are consistent with the primary fit (though the precision within each bin is constrained by the number of Cepheids and SNe Ia available, indicated for each bin). Subsequently, we conduct a joint fit across all bins simultaneously, as illustrated in Figure~\ref{fig:LP_bin}, encompassing ten $\logp$ period bins spanning the range $[1,2]$. This fit assumes only a local power-law $P-L$ relation within each $0.1$-wide $\logp$ bin and remains unaffected by potential period distribution disparities between nearby and faraway galaxies. The resultant determination of $H_0=72.50\pm1.69$ fully aligns with our primary fit outcome (case 2 in Table~\ref{tbl:H0}), with no compelling evidence for a substantially better fit.

In conclusion, our primary $W_H$ fit enables the determination of $H_0$ with an accuracy of approximately $2.3\%$, with a significantly reduced impact from systematic uncertainties.


\section{$W_I$ fitting results}
\label{sec:WI results}

\begin{figure}
\includegraphics[width=0.48\textwidth]{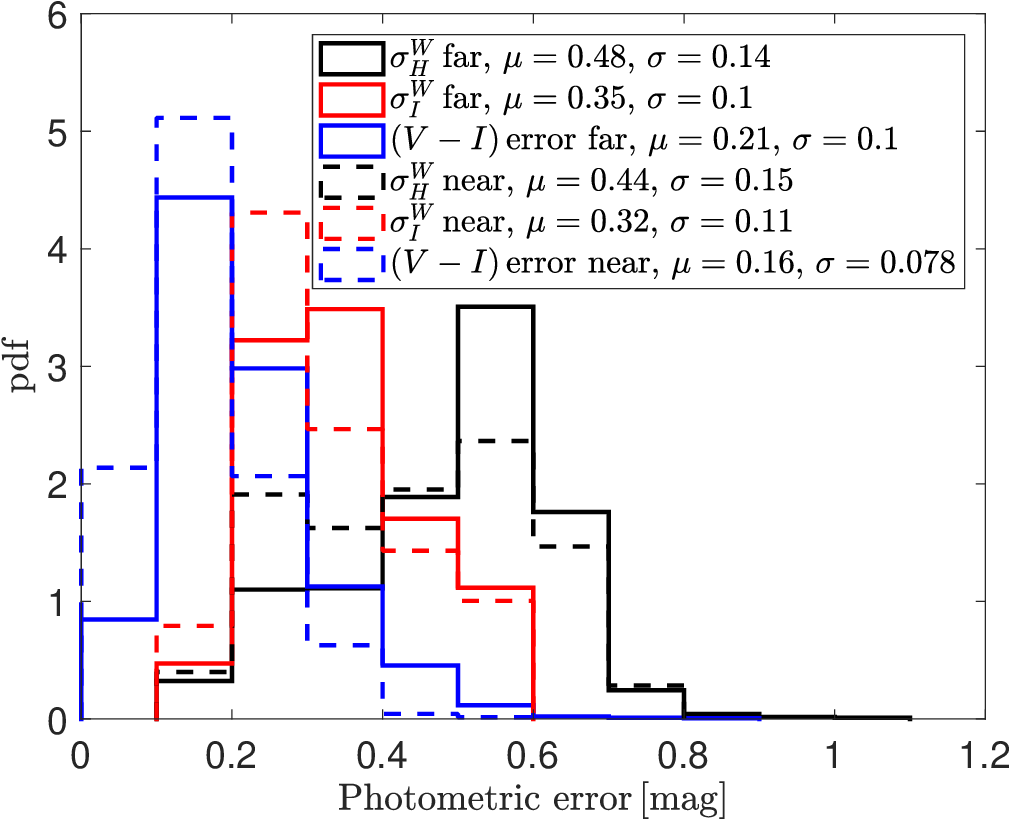}
\caption{The histograms depict the distributions of the total error in magnitudes and colors provided by the R22 data release. In each histogram set, black represents $\sigma_{H}^{W}$, red represents $\sigma_{I}^{W}$, and blue represents the error of the color term (F555W-F814W). Solid histograms correspond to faraway galaxies, while dashed histograms correspond to nearby galaxies. The legend provides the means and standard deviations of the distributions (in magnitudes). We applied a correction of $15\%$, as recommended by the SH0ES collaboration due to the correlation between F555W and F814W background errors, to both $\sigma_{I}^{W}$ and (F555W-F814W). Errors in faraway galaxies are slightly larger than those in nearby galaxies. $\sigma_{I}^{W}$ is smaller than $\sigma_{H}^{W}$, although the extent of this effect is not as substantial as expected.
\label{fig:error}}
\end{figure}

\vspace{20mm}
The minimal systematic uncertainties associated with extinction, metallicity, and the $P-L$ relation shape, as demonstrated for our primary $W_H$ fit (Section~\ref{sec:WH results}), suggest that comparable accuracy can also be achieved with $W_I$. In this section, we validate this assertion. Leveraging optical photometry offers the potential for enhanced precision compared to NIR photometry, given the significantly lower optical background attributed to higher resolution, smaller pixels, and diminished flux from red giants (R22). While $\sigma_{I}^{W}$ is indeed smaller than $\sigma_{H}^{W}$ (refer to Figure~\ref{fig:error}), the magnitude of this effect is not as substantial as anticipated (with the mean of the error distribution being approximately $0.45\,\textrm{mag}$ for $W_H$ and $0.35\,\textrm{mag}$ for $W_I$). Further discussion on the fidelity of the R22-$W_I$ dataset is provided towards the end of this section.

For our primary R22-$W_I$ fit, we adopt $R_I^{VI}=1.18$ \citep[corresponding to $R_V=3.3$ of][which closely aligns with $R_I^{VI}=1.19$ of R22, see Section~\ref{sec:R_Extinction}]{Fitzpatrick1999}, yielding $H_0=76.16\pm1.69$ (case 3 in Table~\ref{tbl:H0}). It is noteworthy that we obtain $\chi_{\nu}^{2}\approx0.901$ for $1299$ degrees of freedom, with a probability of $\myapprox5\times10^{-3}$ to achieve a lower value of $\chi_{\nu}^{2}$. Using the initially published $\sigma_{I,i,j}^{W}$ without the $0.85$ multiplier (see Section~\ref{sec:data}) resulted in $\chi_{\nu}^{2}\approx0.750$ for $1321$ degrees of freedom, with an insignificant chance of obtaining such a low value of $\chi_{\nu}^{2}$\footnote{After we communicated this discrepancy to the SH0ES collaboration, they introduced the $15\%$ correction.}. This relatively low value of $\chi_{\nu}^{2}$ might be attributed to an overestimate of $\sigma_{I,i,j}^{W}$ due to the correlation between F555W and F814W background errors\footnote{SH0ES collaboration, private communication.}, which is presently estimated at $\myapprox15\%$ (and accounted for in our analysis), although not explicitly discussed in any SH0ES publication. If we increase the assumed error overestimation to $20\%$, we obtain $\chi_{\nu}^{2}\approx0.97$ that aligns perfectly with $1291$ degrees of freedom, without altering any of our conclusions below. Consequently, we proceed with the recommended $15\%$ error overestimation for the optical data, mindful that further quality checks may be necessary before achieving the same level of reliability as the NIR data.

The obtained $\myapprox2.2\%$ error in $H_0$ can be interpreted similarly to how it was estimated for the primary $W_H$ fit. We find $\delta\Delta\mu_{\rm{N}4258}\approx0.019\,\rm{mag}$, which can be approximated as $\delta\Delta\mu_{\rm{N}4258}\approx0.33/\sqrt{220}\approx0.022\,\rm{mag}$, where the mean statistical uncertainty of the 220 N4258 Cepheids' Wesenheit magnitude is $\myapprox0.33\,\rm{mag}$. Similarly, we derive $\delta(\Delta\mu_{\rm{host}}+m_B)\approx0.027\,\rm{mag}$, with SNe Ia dominates the error budget, where $\delta_{m_B}\approx0.024\,\rm{mag}$, which can be approximated as $\delta_{m_B}\approx0.11/\sqrt{19}\approx0.025\,\rm{mag}$. Although the photometry of the Cepheids offers higher accuracy, reducing the error of the relative distance between NGC 4258 and the host galaxies based on the Cepheids $P-L$ relation, the smaller number of SNe Ia increases the error, resulting in a total error similar to the $W_H$ case. Our estimate yields $\delta M_B\approx0.046\,\rm{mag}$, leading to an error of $\myapprox2.1\%$ in $H_0$, consistent with the fit results. It is expected that the $H_0$ error with the complete 42 SNe Ia sample of R22 would be $\myapprox1.9\%$, although inclusion of the full covariance matrix for the optical Wesenheit magnitude (not accounted for in our analysis) might slightly alter this estimate.

We proceed to assess the systematic error in determining $H_0$ with our primary R22-$W_I$ fit. Most of the examined variants exhibit negligible effects on $H_0$ ($|\delta H_0|<0.55$). These include incorporating Cepheids with $\logp<1$, with or without a different $P-L$ slope (the impact is minor due to the limited number of Cepheids, $\myapprox200$, with $\logp<1$), exclusion of the outlier rejection, adjustment of the outlier rejection to eliminate the single largest outlier at a time, raising the outlier rejection threshold to $3.5\sigma$ (with either global or single-outlier-at-a-time rejection), implementing a period cutoff $\logp<1.6$, disregarding the metallicity term (i.e., setting $Z_I^{W}=0$), adopting $R_I^{VI}=0.95$ \citep[smaller than $R_V=2.5$ of][]{Fitzpatrick1999}, adopting $R_I^{VI}=1.4$ \citep[larger than $R_V=3.3$ of][]{Cardelli1989}, and including M31 Cepheids (with minimal impact due to the limited number of $\myapprox50$ Cepheids in M31 with $\logp>1$). The sole variant exhibiting a non-negligible effect on the value of $H_0$ is the adjustment of the period cutoff to $\logp>1.2$ ($\delta H_0\approx-1.26$). Given the marginal effects of all fit variants, we conclude that the systematic uncertainty can be disregarded in this case as well. One particularly surprising outcome is the observed low sensitivity to $R_I^{VI}$, considering the conventional rationale for employing $W_H$ due to the presumed high sensitivity of $W_I$ to extinction correction. We elucidate the reason for this diminished sensitivity with our sample in the subsequent section.

\begin{figure}
\includegraphics[width=0.48\textwidth]{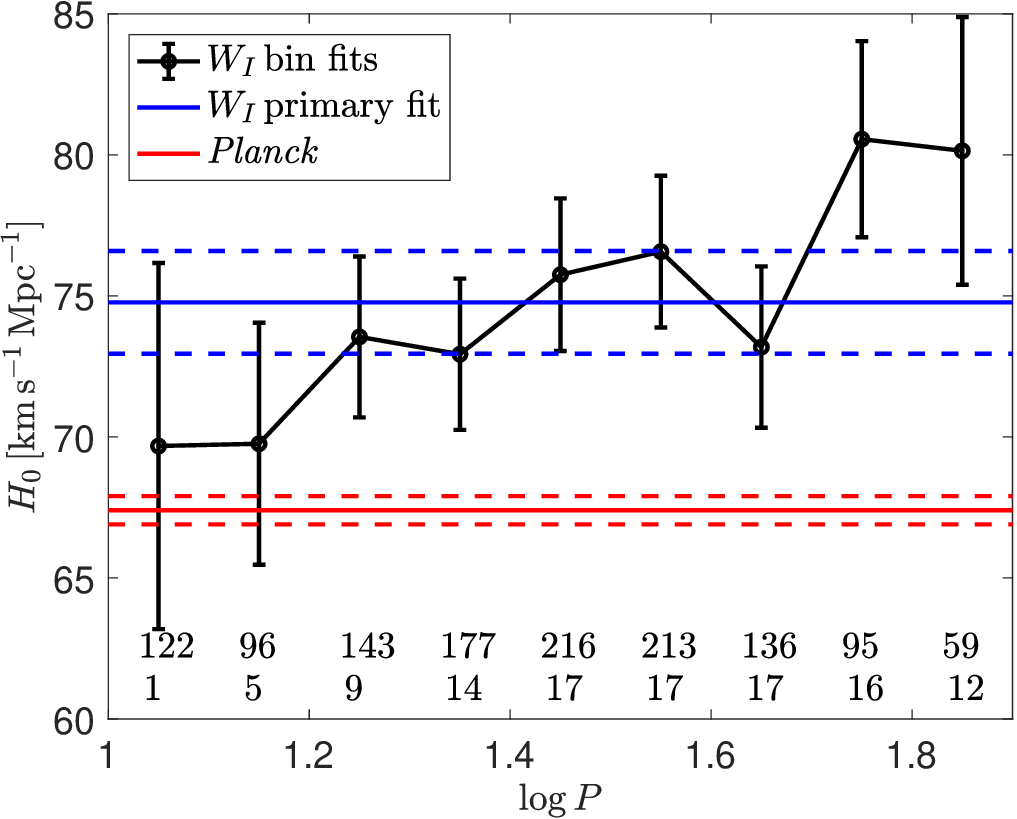}
\caption{Results of fitting using $W_I$ in $0.1-\logp$ bins. The primary fit result (blue, solid line represents the best fit, and dashed lines indicate one standard deviation) aligns with the fit in each bin (though the accuracy in each period bin is restricted by the number of Cepheids and SNe Ia, indicated for each bin). An apparent trend in the fit outcomes (black; errors denote one standard deviation) with period is observable. By fitting $H_0$ in each bin as a linear function of $\logp$, we derive a slope of $11.4\pm3.5\,\rm{km}\,\rm{s}^{-1}\,\rm{Mpc}^{-1}\,\rm{dex}^{-1}$ (a significance of $\myapprox3.3\sigma$). This should be considered an upper limit to the trend, as we have not accounted for the correlation among some bins stemming from shared hosts, and all bins are influenced by the NGC 4258 distance error. Fitting only for $\logp>1.2$, where each bins contains a significant fraction of the hosts (to minimize the variance of the SNe Ia sample compared with the Cepheid sample), we obtain a slope of $10.1\pm4.7\,\rm{km}\,\rm{s}^{-1}\,\rm{Mpc}^{-1}\,\rm{dex}^{-1}$ (a significance of $\myapprox2.2\sigma$). Red lines denote the \textit{Planck} results.
\label{fig:Bin_Plot_WI}}
\end{figure}

We next replicate our binned analysis for the $W_I$ scenario. The outcomes of the fits in $0.1-\logp$ bins are displayed in Figure~\ref{fig:Bin_Plot_WI}. As depicted in the figure, although the fit outcomes in each bin align with the primary fit result (albeit with accuracy limited by the number of Cepheids and SNe Ia in each period bin, indicated in the figure), there appears to be a discernible trend with period in this case. Upon fitting $H_0$ in each bin as a linear function of $\logp$, we ascertain a slope of $11.4\pm3.5\,\rm{km}\,\rm{s}^{-1}\,\rm{Mpc}^{-1}\,\rm{dex}^{-1}$ (a significance of $\myapprox3.3\sigma$). This should be considered an upper limit to the trend, as we have not accounted for the correlation among some bins stemming from shared hosts, and all bins are influenced by the NGC 4258 distance error. Fitting only for $\logp>1.2$, where each bins contains a significant fraction of the hosts (to minimize the variance of the SNe Ia sample compared with the Cepheid sample), we obtain a slope of $10.1\pm4.7\,\rm{km}\,\rm{s}^{-1}\,\rm{Mpc}^{-1}\,\rm{dex}^{-1}$ (a significance of $\myapprox2.2\sigma$). This observation might signify additional concerns with the optical photometry (in addition to the overestimated $\sigma_{I,i,j}^{W}$). Subsequently, upon fitting all bins collectively, we arrive at $H_0=75.25\pm1.71$, aligning with the primary fit result (case 4 in Table~\ref{tbl:H0}). We discern no substantial evidence for a superior fit, with a $\chi^2$ smaller by approximately 59 from a single, $[1,2]$, $\logp$-bin fit and 37 fewer degrees of freedom.

\section{Limited Impact of Extinction Corrections}
\label{sec:Low extinction}

In the preceding section, we illustrated the minimal impact of extinction correction on the $W_I$ fitting results. Here, we delve into the rationale behind this observation. We commence with a discussion on extinction corrections (subsection~\ref{sec:R_Extinction}), followed by an evaluation of the extinction distribution of extragalactic Cepheids (subsection~\ref{sec:distribution}). Subsequently, in subsection~\ref{sec:MW}, we showcase a substantial sensitivity to extinction when incorporating MW Cepheids.

\subsection{Correction for extinction}
\label{sec:R_Extinction}

The rationale behind employing the Wesenheit index lies in its minimal susceptibility to extinction \citep{Madore1982}. Defined for three bands,  $X$, $Y$, and $Z$, the index is expressed as
\begin{eqnarray}\label{eq:Weseheit general}
W=X-R_{X}^{YZ}\left(Y-Z\right),
\end{eqnarray}
measured in magnitudes. In the presence of extinction, this can be expressed as
\begin{eqnarray}\label{eq:W=W0}
W&=&(X-X_0)+X_0-R_{X}^{YZ}E(Y-Z)-R_{X}^{YZ}\left(Y-Z\right)_0\nonumber\\
&=&X_0-R_{X}^{YZ}\left(Y-Z\right)_0\equiv W_0,
\end{eqnarray}   
where the subscript $0$ represents the extinction-free magnitude, $E(Y-Z)=(Y-Z)-(Y-Z)_0$ denotes the selective extinction, and the final equality holds if $R_{X}^{YZ}=(X-X_0)/E(Y-Z)=A_X/E(Y-Z)\equiv \tilde{R}^{X}_{YZ}$, with $A_X$ representing the total extinction in band $X$. An often-overlooked aspect is that the Wesenheit index doesn't correct $X$ for extinction; instead, with the appropriate choice of $R_{X}^{YZ}$, it yields a quantity insensitive to extinction (also noted in R22). For any arbitrary value of $R_{X}^{YZ}$, the deviation of $W$ from $W_0$ is given by
\begin{eqnarray}\label{eq:W-W0}
W-W_0&=&A_X-R_{X}^{YZ}E(Y-Z)\nonumber\\
&=&\left(\tilde{R}^{X}_{YZ}-R_{X}^{YZ}\right)E(Y-Z),
\end{eqnarray}  
such that even if $\tilde{R}^{X}_{YZ}$ is not precisely known or varies across different lines-of-sight towards the Cepheids, the deviation in $W-W_0$ remains minimal as long as the selective extinction $E(Y-Z)$ remains small. With zero extinction, $E(Y-Z)=0$, any value of $R_{X}^{YZ}$ can be chosen. However, this value must be consistent across all Cepheids; otherwise, disparate quantities are being compared. \citet{Mortsell2021a,Perivolaropoulos2021} selected different values of $R_X^{YZ}$ for different Cepheids, rendering their conclusion invalid for that analysis (also highlighted in R22).

An alternative approach involves directly adjusting $X$ for extinction \citep{Follin2018}:
\begin{eqnarray}\label{eq:RE}
F&=&X-R_{X}^{YZ}E(Y-Z),\nonumber\\
&=&(X-X_0)+X_0-R_{X}^{YZ}\left[(Y-Z)-(Y-Z)_0\right]\nonumber\\
&=&X_0,
\end{eqnarray} 
where the last equality holds if $R_{X}^{YZ}=\tilde{R}^{X}_{YZ}$. Similar to the $W$ case, for any given value of  $R_{X}^{YZ}$, the deviation of $F$ from $X_0\equiv F_0$ is given by
\begin{eqnarray}\label{eq:X-X0}
F-F_0&=\left(\tilde{R}^{X}_{YZ}-R_{X}^{YZ}\right)E(Y-Z).
\end{eqnarray}  
The distinction lies in the allowance for choosing different $R_{X}^{YZ}$ values for different Cepheids, as the color term vanishes under zero extinction. However, this method assumes knowledge of $(Y-Z)_0$, which is not always available, such as for $(V-I)_0$ in the range $\logp>1.72$ \citep{Sharon2022}, where a significant fraction of the extragalactic Cepheids is situated (see Figure~\ref{fig:Count}). It is worth noting that if $(Y-Z)_0$ follows a linear relationship with $\logp$, the analysis with $F$ is identical to that with $W$ (for a single global $R_{X}^{YZ}$ value), as the difference between $E(Y-Z)$ and $(Y-Z)$ gets absorbed into $M_{k}^{W}$ and $b_k^{W}$ \citep{Follin2018}. 

We infer that minimizing sensitivity to extinction corrections involves selecting Cepheids with minimal extinction. As demonstrated in Section~\ref{sec:distribution}, this holds true for both nearby and faraway galaxies, elucidating the reduced susceptibility of our dataset to extinction correction.

\subsection{The extinction distribution}
\label{sec:distribution}

\begin{figure}
\includegraphics[width=0.48\textwidth]{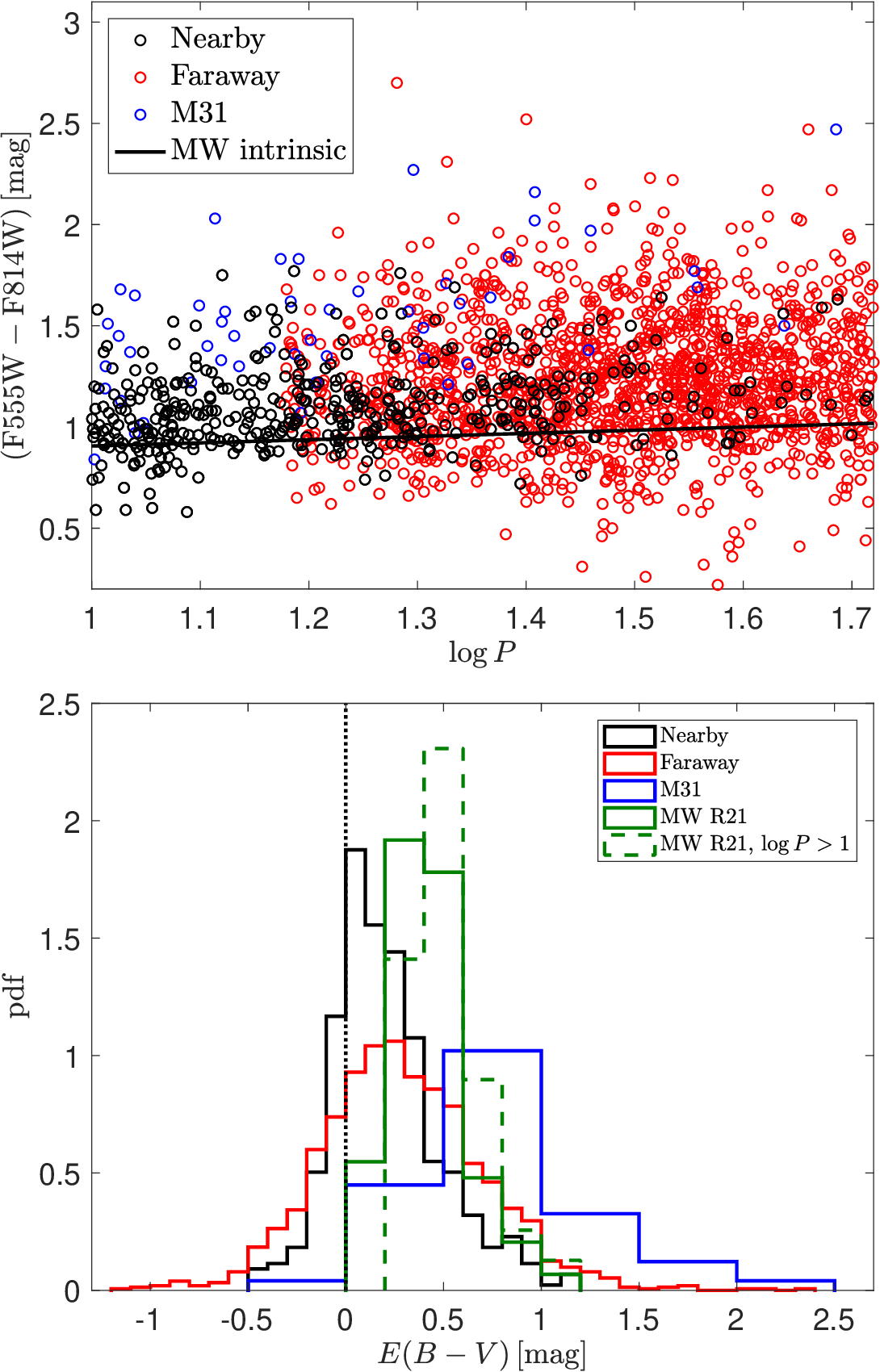}
\caption{Top panel: the $P-C$ relation for the (F555W-F814W) color of the R22-$W_{H}$ sample. Nearby (faraway) Cepheids from M101 and NGC 4258 (the other hosts) are denoted by black (red) circles, while M31 Cepheids are represented by blue circles. The color errors (not depicted) are primarily influenced by dispersion due to blending, with values of approximately $\myapprox0.15\,\rm{mag}$ for the nearby sample and $\myapprox0.2\,\rm{mag}$ for the faraway sample. The intrinsic MW $P-C$ relation, Equation~\eqref{eq:intrinsic MW}, is illustrated by the black line. Bottom panel: The distributions of the host $E(B-V)$, derived using Equation~\eqref{eq:E(B-V)} and corrected for (the small, $\mylesssim0.1$\,\rm{mag}) Galactic extinction \citep{Schlafly2011}. Nearby (faraway) galaxies are represented by solid black (red) lines, while M31 Cepheids are shown in blue. The solid (dashed) green lines denote MW Cepheids from the sample of \citet{Riess2021a}, with MW $E(B-V)$ values sourced from the catalog of \citet{Sharon2022}, without (with) a $\logp>1$ restriction. The $E(B-V)$ distributions of the nearby sample exhibit a pronounced change at $E(B-V)\approx0$, indicative of a blue edge. The small number of Cepheids with negative $E(B-V)$ values (physically implausible) are in line with the scatter of $\sigma_{\rm{int}}\approx0.1\,\rm{mag}$ around the intrinsic $P-C$ relation (see Appendix~\ref{sec:P-C}), as well as the color error attributed to blending. The blue edge of the faraway sample's $E(B-V)$ distribution appears somewhat less distinct, likely due to larger color errors in faraway galaxies that can obscure the blue edge. Cepheids in nearby and faraway galaxies experience relatively low extinction, while M31 and MW Cepheids are subject to more substantial extinction.
\label{fig:R16_colors}}
\end{figure}

We initiate the analysis by assessing the reddening distribution of extragalactic Cepheids (utilizing the R22-$W_{H}$ sample unless specified otherwise). The top panel of Figure~\ref{fig:R16_colors} displays the (F555W-F814W) color of extragalactic Cepheids against their period. We confine this examination to $\logp<1.72$, where an accurate determination of the intrinsic MW Cepheids colors is available \citep{Sharon2022}. In Appendix~\ref{sec:P-C}, we compute the intrinsic color for $1<\logp<1.72$ \citep[employing methods from][]{Tammann2003}:
\begin{eqnarray}\label{eq:intrinsic MW}
(\textrm{F555W-F814W})_{0}&\approx&0.156\logp+0.75,
\end{eqnarray}
represented by a solid black line. Additionally, we distinguish between nearby (M101 and NGC 4258; black circles) and faraway (the other hosts; red circles) galaxies, considering that Cepheids in nearby galaxies are measured with slightly higher precision (see Figure~\ref{fig:error}). The errors in (F555W-F814W) color (not depicted) primarily stem from dispersion due to blending, amounting to approximately $0.15,\rm{mag}$ for the nearby sample and $0.2,\rm{mag}$ for the faraway sample.

Subsequently, we estimate the host galaxy $E(B-V)$ toward each Cepheid using (see Appendix~\ref{sec:P-C}):
\begin{eqnarray}\label{eq:E(B-V)}
E(B-V)&\approx&1.42[(\textrm{F555W-F814W})-(\textrm{F555W-F814W})_{0}],\nonumber\\
\end{eqnarray}
and then we correct for (the small, $\mylesssim0.1$\,\rm{mag}) Galactic extinction \citep{Schlafly2011}. The resulting $E(B-V)$ distribution is depicted in the bottom panel of Figure~\ref{fig:R16_colors}. As illustrated in the figure, the $E(B-V)$ distribution of the nearby sample (solid black line) exhibits a sharp change at $E(B-V)\approx0$, indicative of a blue edge. This blue edge corroborates both Equation~\eqref{eq:intrinsic MW} and the photometry of nearby Cepheids in (F555W-F814W), which were determined through entirely independent methods\footnote{For instance, deriving Equation~\eqref{eq:intrinsic MW} necessitates knowledge of the extinction toward MW Cepheids, sourced from \citet{Fernie1995,Turner2016,Groenewegen2020}.}. The small number of Cepheids with negative $E(B-V)$ (physically implausible values) aligns with the scatter of $\sigma_{\rm{int}}\approx0.1\,\rm{mag}$ around the intrinsic $P-C$ relation \citep[refer to Appendix~\ref{sec:P-C}, corresponding to the width of the instability strip;][]{Tammann2003}, as well as with color errors stemming from blending. The blue edge of the faraway sample's $E(B-V)$ distribution (solid red line) appears somewhat less distinct, which is understandable given the larger color errors associated with faraway galaxies that can blur the blue edge. 

The distributions' shapes are influenced by various selection biases, notably favoring Cepheids with minimal extinction ($E(B-V)\lesssim0.5$), resulting in a negligible impact of extinction corrections. To further illustrate this minimal impact, we perform our primary fit without the color term, setting $R_I^{VI}=R_H^{VI}=0$. For $W_H$, we derive $H_0=72.39\pm1.66$ (case 5 in Table \ref{tbl:H0}), compared to $H_0=72.68\pm1.67$ with $R_H^{VI}=0.386$. Similarly, for $W_I$, we obtain $H_0=74.42\pm1.66$ (case 6 in Table \ref{tbl:H0}), contrasting with $H_0=76.16\pm1.69$ with $R_I^{VI}=1.18$. The overall effect on $H_0$ amounts to $\delta H_0\approx0.3$ for $W_H$ and $\delta H_0\approx1.7$ for $W_I$, validating our assertion of small systematic uncertainty in Sections~\ref{sec:WH results} and~\ref{sec:WI results}.

\subsection{Incorporating MW Cepheids}
\label{sec:MW}

Here, we illustrate the significant sensitivity of the $W_I$ fitting results to extinction that arises when including MW Cepheids. Initially, we repeat our primary $W_H$ fit with the inclusion of MW Cepheids, employing the MW anchor instead of the NGC 4258 anchor. Following the methodology of \citet{Mortsell2021a}, we transform the residual Gaia parallax calibration offset, $zp$, into a linear parameter and fit for it. The data for MW Cepheids are sourced from Table 1 in \citet{Riess2021a}. Since a significant fraction of the MW Cepheids have $\logp<1$, we do not restrict the Cepheid sample to $\logp>1$ in this case (for all galaxies). We obtain $H_0=74.43\pm2.21$\footnote{We incorporate an intrinsic dispersion of $\sigma_{\rm{int}}=0.069\,\rm{mag}$ for $W_H$ of the MW Cepheids \citep{Riess2019a}.} (case 7 in Table~\ref{tbl:H0}), consistent with the result using the NGC 4258 anchor, $H_0=72.68\pm1.67$. Similarly small impact is observed for our primary $W_I$ fit\footnote{Here, we adopt $\sigma_{\rm{int}}=0.085\,\rm{mag}$ for the intrinsic dispersion \citep[upper limit from Table 3 of][]{Riess2019a}.} ($H_0=75.67\pm2.24$ for the MW anchor, case 8 in Table~\ref{tbl:H0}, compared to $H_0=76.16\pm1.69$ for the NGC 4258 anchor). Subsequently, we demonstrate the considerable systematic uncertainty of the $W_I$ fitting results related to the extinction law, attributed to the substantial extinction associated with MW Cepheids.

The bottom panel of Figure~\ref{fig:R16_colors} illustrates the selective extinction affecting the MW Cepheids utilized in our analysis. These data are sourced from the catalog of \citet{Sharon2022}, with detailed references therein outlining the methods employed to estimate the extinction. As depicted in the figure, MW Cepheids, particularly those with $\logp>1$, endure substantial extinction due to their location within the Galactic disc, where massive stars are prevalent, relative to the position of the Sun. When using the MW anchor instead of the NGC 4258 anchor and setting $R_I^{VI}=R_H^{VI}=0$, we obtain $H_0=74.28\pm2.21$ for $W_H$ (case 9 in Table \ref{tbl:H0}) and $H_0=83.1\pm2.6$ for $W_I$ (case 10 in Table \ref{tbl:H0}). While a negligible effect is obtained for the $W_H$ case, the effect is notably pronounced in the $W_I$ case ($\delta H_0\approx7.5$), indicating that even minor uncertainties in correcting for this effect would result in significant systematic uncertainties. The heightened effect observed for $W_I$ primarily motivates observations with the F160W filter. However, as demonstrated in Section~\ref{sec:distribution}, this effect is considerably mitigated for our sample. 

A comparable scenario arises for M31. The selective extinction affecting M31 Cepheids is depicted in the bottom panel of Figure~\ref{fig:R16_colors}. This distribution is derived using the methodologies outlined in Section~\ref{sec:distribution}. As illustrated in the figure, M31 Cepheids experience substantial extinction due to the nearly edge-on orientation of M31's disk, but close enough to allows for the measurement of obscured Cepheids.


\section{Discussion and Summary}
\label{sec:summary}

The SH0ES collaboration employs a comprehensive approach to ascertain $H_0$, combining host Cepheids with a diverse array of Cepheids from various observations. This strategy minimizes statistical uncertainty in $H_0$ and harnesses multiple independent geometrical distance anchors, mitigating sensitivity to potential systematic errors in any single anchor distance determination. Cepheids from NGC 4258, as well as from closer galaxies like the MW, LMC, SMC, and M31, are incorporated to allow the use of multiple anchors and to enhance the precision of the $P-L$ relation. However, this Cepheid sample with a wide-distance range is susceptible to systematic errors due to the comparison of distant host Cepheids with those much closer. Host and NGC 4258 Cepheids are measured via HST imaging, while closer Cepheids are identified through archival observations and observed using different techniques. Corrections are applied to account for differences in observing methodologies and biases in period distributions and extinction levels between close and host Cepheids. Additionally, the metallicity discrepancy between LMC/SMC and host Cepheids is addressed by considering a metallicity-dependent $P-L$ relation.

We proposed utilizing a subset of SH0ES Cepheids to derive $H_0$ independently of potential systematic errors associated with Cepheids. This subset excludes Cepheids from the MW, LMC, SMC, and M31, retaining only the water megamasers in NGC 4258 as a single anchor. By omitting these Cepheids, which are subject to different observational methods and reduction techniques compared to those in host galaxies and NGC 4258, potential systematic errors in Cepheid measurement are circumvented, encompassing identification, selection, period determination, photometry, template usage, and crowding correction. We demonstrate the elimination or substantial reduction of uncertainties pertaining to reddening law, metallicity sensitivity, and breaks in the $P-L$ relation, facilitating precise determination of Cepheid relative distances. Remaining statistical errors in Cepheid relative distance determination, primarily associated with Cepheid and SNe Ia brightness zero points, are sufficiently small for competitiveness, $\myapprox2.3\%$ (or about 1.65 km/s/Mpc) using the R22 data release. The Cepheid-related systematic errors are significantly smaller than the statistical uncertainty, rendering the accuracy adequate to identify $\myapprox3\sigma$ tension with the \textit{Planck} $H_0$ value. 

Once the decision is made to forego the local anchor points, there is no inherent requirement to shift to the NIR. As demonstrated in Section~\ref{sec:WI results}, we established that the systematic uncertainty associated with optical photometry is considerably smaller than the statistical uncertainty. This implies the potential for achieving greater accuracy compared to NIR photometry, primarily because the optical background is approximately an order of magnitude lower than that in the NIR, owing to factors such as higher resolution, smaller pixels, and reduced flux from red giants (R22). The observed low sensitivity to extinction correction stems from the minimal extinction experienced by Cepheids within our sample (Section~\ref{sec:Low extinction}). However, in practice, the quality of photometry in optical bands does not exhibit substantial improvement over NIR photometry, and the reliability of the optical photometry sample does not match that of the NIR photometry sample, as demonstrated in Section~\ref{sec:WI results}.

The primary drawback of our approach lies in the pronounced reliance of $H_0$ on the distance to NGC 4258, $D_{4258}$, as expressed by the equation:
\begin{eqnarray}\label{eq:D4258}
\frac{dH_0}{H_0}\approx-f\frac{dD_{425}}{D_{4258}},
\end{eqnarray} 
where $f\approx1$, in contrast to $f\approx1/3$ with the inclusion of three anchors (NGC 4258, LMC, and MW). This heightened dependency renders our method more susceptible to systematic uncertainties in $D_{4258}$. Nonetheless, given the resilience of our method against Cepheid-related systematic errors and the absence of discernible signs of SNe Ia-related systematic errors in detailed examination  \citep[R22,][but see also \citealt{Steinhardt2020,Wojtak2022,Wojtak2024}]{Carr2022,Brout2022,Peterson2022}, our analysis suggests two potential scenarios: either standard cosmology requires revision or there exists a systematic error in the current determination of the distance to NGC 4258, $D_{4258}\approx7.57\pm0.24\,\textrm{Mpc}$. To fully alleviate the Hubble tension, a distance of $D_{4258}\approx8.15\,\textrm{Mpc}$ is necessary, and considering the prevailing uncertainties, a distance (with a central value) exceeding $D_{4258}\gtrsim7.8\,\textrm{Mpc}$ is needed to reduce the tension below $2\sigma$. Moreover, the scenario of systematic error in the distance to NGC 4258 to resolve the Hubble tension would also necessitate an unrelated systematic error in the comparison of distant host Cepheids with those much closer to explain the $H_0$ measurement of R22. However, solid evidence supporting this assertion has not been presented so far. 

Our investigation underscores the importance of conducting a meticulous analysis of optical observations of extragalactic Cepheids to enable the application of our method with an optical Cepheid sample. Looking ahead, observations with JWST hold promise for further reducing optical crowding corrections \citep{Riess2024}. Moreover, the heightened sensitivity of JWST may facilitate the measurement of Cepheids in additional megamaser galaxies with precise geometrical distances, such as UGC 3789 \citep[$D=49.6\pm5.1\,\rm{Mpc}$;][]{Reid2013}, thereby enabling the extension of our method with supplementary anchors.

\section*{Acknowledgements}
We thank members of the SH0ES collaboration: Adam Riess, Dan Scolnic, Lucas Macri, Stefano Casertano, and Wenlong Yuan for extensive correspondences and discussions that significantly improved the manuscript. We thank Andy Gould, Eli Waxman, Boaz Katz, Barak Zackay, Subo Dong, and Eran Ofek for useful discussions. DK is supported by a research grant from The Abramson Family Center for Young Scientists, and by the Minerva Stiftung. 

\section*{Data availability}

All data used in this study are publicly available through other publications. 






\begin{appendix}

\section{The period-color relation}
\label{sec:P-C}

\begin{figure*}
	\includegraphics[width=1\textwidth]{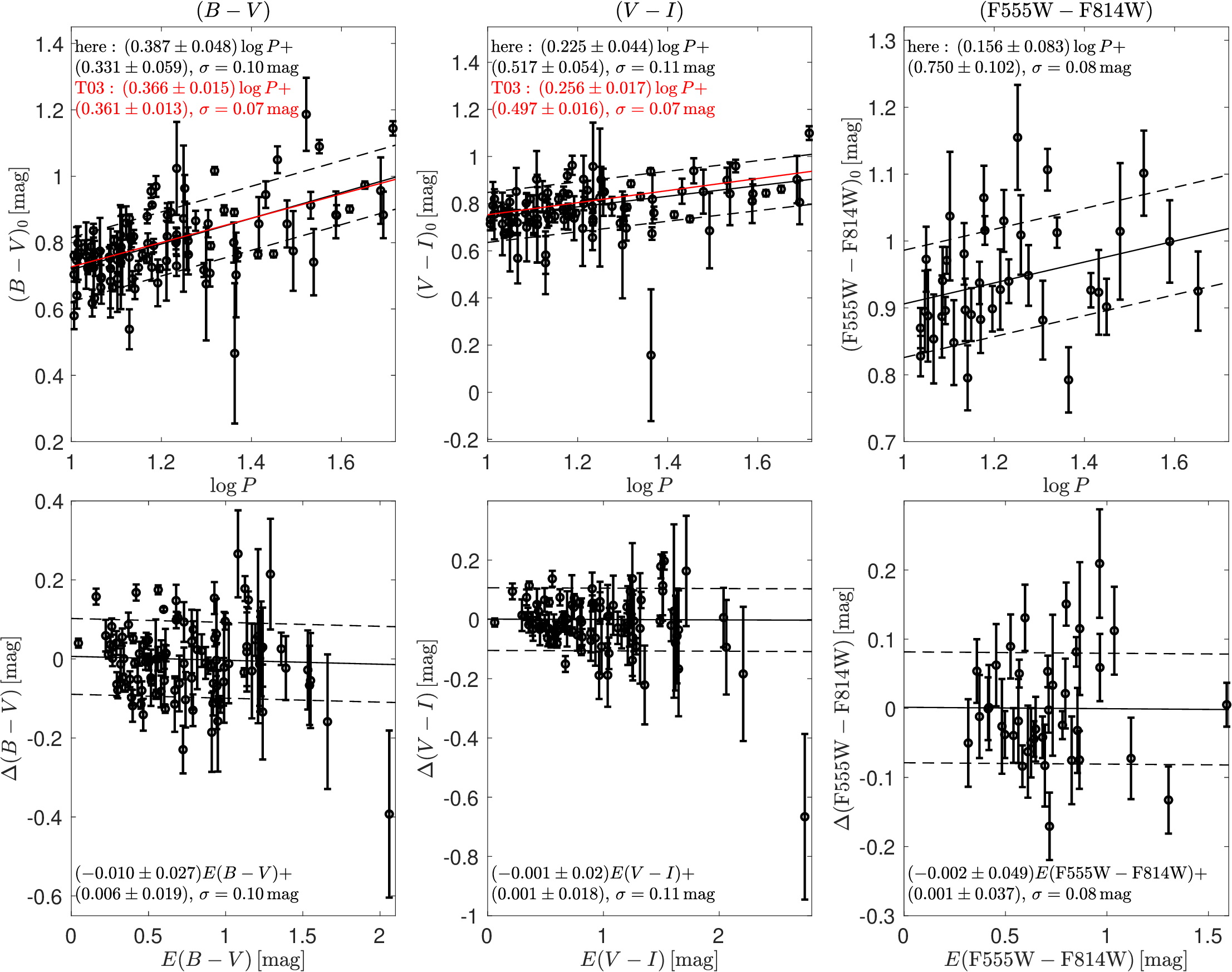}
	\caption{Determining the intrinsic $P-C$ relation of Cepheids with $1<\log P<1.72$. The upper left panel displays the direct determination $(B-V)_0=(B-V)-E(B-V)$ (black symbols). Fitting $(B-V)_0$ as a linear function to $\log P$ yields a large reduced $\chi^2$, indicating intrinsic scatter around the fit (with a calibrated intrinsic scatter of $\myapprox0.07\,\rm{mag}$). Incorporating this calibrated intrinsic scatter, the resulting fit is $(B-V)_0=(0.387\pm0.048)\logp+(0.331\pm0.059)\,\rm{mag}$ (black lines, with a scatter of $0.10\,\rm{mag}$). In the bottom left panel, we present the distance of each Cepheid from the best-fit line, $\Delta(B-V)$, plotted against the estimated $E(B-V)$. No trend is observed (the slope is $-0.010\pm0.027$, black lines), indicating no systematic scale error in the determined $E(B-V)$ values. Our derived intrinsic $(B-V)_0$ color closely matches the result of \citet[][red line]{Tammann2003}, albeit utilizing updated $E(B-V)$ values from \citet{Turner2016} for $56$ Cepheids and restricting Cepheid periods to $1<\log P<1.72$. Middle panels: We calibrate $E(V-I)=1.33E(B-V)$, and repeating the analysis of $(B-V)$ for $(V-I)$ reveals no dependence of $\Delta(V-I)$ on $E(V-I)$. The best fit we obtain is $(V-I)_0=(0.225\pm0.044)\logp+(0.517\pm0.054)\,\rm{mag}$ (black lines, with a scatter of $0.11\,\rm{mag}$), and the ratio between $E(V-I)$ and $E(B-V)$ closely resembles the findings of \citet[][red line]{Tammann2003}. Right panels: Repeating the procedure for $(\textrm{F555W}-\textrm{F814W})$, we calibrate $E(\textrm{F555W}-\textrm{F814W})=1.42E(B-V)$. The resulting fit is  $(\textrm{F555W}-\textrm{F814W})_0=(0.156\pm0.083)\logp+(0.750\pm0.102)\,\rm{mag}$ (black lines, with a scatter of $0.08\,\rm{mag}$).}
	\label{fig:PC}
\end{figure*}

This appendix utilizes the catalog provided by \citet{Sharon2022} to investigate the intrinsic period-color ($P-C$) relationship of Cepheids falling within the range of $1 < \log P < 1.72$ across various colors, as pertinent to the analysis discussed in Section~\ref{sec:Low extinction}. We adopt the methodology outlined by \citet{Tammann2003}. Initially, we compute $(B-V)_0$ by subtracting $E(B-V)$\footnote{$E(B-V)$ values in the catalog are sourced from \citet{Fernie1995,Turner2016,Groenewegen2020}.} from $(B-V)$, as illustrated in the upper left panel of Figure~\ref{fig:PC} (following the exclusion of outliers VZ-Pup and ER-Aur). We fit $(B-V)_0$ against $\log P$ linearly and observe a relatively high reduced $\chi^2$, suggesting that the scatter around the fit is intrinsic, so we calibrate an intrinsic scatter of approximately $0.07$ mag to obtain a unity for the reduced $\chi^2$. Upon incorporating the calibrated intrinsic scatter, the resulting fit is $(B-V)_0=(0.387\pm0.048)\logp+(0.331\pm0.059)\,\rm{mag}$ (depicted by black lines, with a scatter of $0.10$ mag). In the bottom left panel, we depict the deviation of each Cepheid from the best-fit line, $\Delta(B-V)$, against the estimated $E(B-V)$. No discernible trend is observed (the slope is $-0.010 \pm 0.027$, indicated by black lines), indicating the absence of a systematic scaling error in the determined $E(B-V)$ values. Notably, our derived intrinsic $(B-V)_0$ color closely resembles the findings of \citet[][illustrated by the red line]{Tammann2003}, albeit with updated $E(B-V)$ values sourced from \citet{Turner2016} for $56$ Cepheids and a limitation of Cepheid periods to $1 < \log P < 1.72$.

We continue to adhere to the methodology outlined by \citet{Tammann2003} for the determination of $(V-I)_0$. By calibrating $E(V-I) = 1.33E(B-V)$, and subsequently analyzing $(V-I)$ in a manner akin to that of $(B-V)$, we find no discernible dependence of $\Delta(V-I)$ on $E(V-I)$ (depicted in the middle panels of Figure~\ref{fig:PC}, following the exclusion of outlier V0396-Cyg). Our best-fit result is $(V-I)_0=(0.225\pm0.044)\logp+(0.517\pm0.054)\,\rm{mag}$ (represented by black lines, with a scatter of $0.11$ mag). Additionally, the ratio between $E(V-I)$ and $E(B-V)$, as well as the obtained best fit, closely resemble the findings of \citet[][highlighted by the red line]{Tammann2003}.

Finally, we calibrate $E(\textrm{F555W}-\textrm{F814W}) = 1.42E(B-V)$ (as depicted in the right panels of Figure~\ref{fig:PC}), resulting in $(\textrm{F555W}-\textrm{F814W})_0=(0.156\pm0.083)\logp+(0.750\pm0.102)\,\rm{mag}$ (illustrated by black lines, with a scatter of $0.08$ mag).

\end{appendix}

\bsp	
\label{lastpage}
\end{document}